\documentclass[pre,aps,onecolumn]{revtex4}
\usepackage{titlesec, blindtext, color, graphicx}

\usepackage[czech]{babel}
\usepackage[IL2]{fontenc}     
\usepackage{amsmath}
\usepackage{amssymb}
\usepackage{graphicx}
\usepackage{color}
\usepackage{bm}
\usepackage{cslatexquotes} 

\definecolor{gray75}{gray}{0.75}
\newcommand{\hsp}{\hspace*{20pt}}
\titleformat{\chapter}[hang]{\Huge\bfseries}{\thechapter\hsp$\textcolor{gray75}\vert$\hsp}{0pt}{\Huge\bfseries}

\newcommand{\bb}{\begin{equation}}
\newcommand{\ee}{\end{equation}}
\newcommand{\ba}{\begin{eqnarray*}}
\newcommand{\ea}{\end{eqnarray*}}

\newcommand{\dd}{{\rm d}}
\newcommand{\rr}{{\mathbf r}}

\begin{document}

\title{Povrchové fázové pøechody a kritické jevy}

\author{Alexandr \surname{Malijevsk\'y}, Praha}

\maketitle

\section{Úvod}

Když se J. C. Maxwell zaèal v pokroèilém vìku uèit holandsky, nebylo to ze snobských, nýbrž ryze pragmatických dùvodù. Dozvìdìl se totiž o (v mateøštinì psané) doktorské
práci mladého van der Waalse \cite{vdw}, která elegantním zpùsobem s pomocí jednoduchého molekulárního modelu vysvìtlovala fázové chování tekutin a existenci kritického
bodu a která sehrála ve vývoji fyziky zlomovou úlohu. Tato molekulární teorie odstartovala spolu s experimenty T. Andrewse \cite{andrews} \emph{klasickou éru} teorie
fázových pøechodù a kritických jevù, na jejichž objevy navázaly práce P. Weisse a P. Curieho zabývajících se studiem magnetù a speciálnì zpùsobu, jakým feromagnety
ztrácejí pøi vyšších teplotách své aktivní magnetické vlastnosti. Z velmi úzké a pøekvapující analogie mezi chováním tekutin  v okolí kritického bodu a feromagnetù v
okolí Curieho bodu se zaèíná již zaèátkem dvacátého století  vynoøovat jeden z nejpozoruhodnìjších principù fyziky vùbec, tzv. princip \emph{univerzality} \cite{kor},
podle nìhož kritické chování mnohých látek je zcela identické bez ohledu na jejich chemickou rùznorodost. Tato úzká podobnost byla vysvìtlena až s formulací jednoduché,
ale nesmírnì užiteèné Landauovy mesoskopické \emph{teorie støedního pole} (pùvodnì uèené pro vysvìtlení supratekutosti), v rámci níž je rozpoznán význam symetrie a
dimenze jako urèujících faktorù pro chování systému v blízkosti kritického bodu. Ústøedním bodem Landauovy teorii je pøedpoklad, že volná energie systému je analytickou
funkcí \emph{parametru uspoøádání}, jehož hodnota slouží k identifikaci dané fáze, a že je tudíž možné volnou energii rozvést v mocniny tohoto parametru tak, aby byla
respektována symetrie problému.

Konec této klasické etapy hlásí rok 1944, kdy L. Onsager nachází pøesné øešení Isingova modelu ve dvou dimenzích. Výsledky této práce byly
zdrcující, nebo odhalily nedostateènost teorie støedního pole a její predikce hodnot kritických koeficentù. Dùvodem nezdaru celé tøídy tìchto
klasických teorií (tj. Landauova typu) je neadekvátní zahrnutí vlivu tepelných fluktuací, jejichž význam roste s redukcí dimenze systému. Naopak, pro
každý kritický jev existuje jistá mezní nebo též horní kritická dimenze $d^*$, nad níž jsou výsledky plynoucí z aproximace støedního pole již pøesné.

Uspokojivé postupy zahrnující v dostateèné míøe fluktuaèní jevy se zaèaly postupnì vyvíjet až v prùbìhu šedesátých a zaèátkem sedmdesátých  let a
byly završeny formulací teorie renormalizaèní grupy  K. Wilsonem \cite{wilson} (inspirované metodami z fyziky vysokých energií a též \clqq blokovou
metodou\crqq $ $ L. Kadanoffa), která mimo jiné vysvìtluje pùvod principu univerzality. Po dovršení této etapy, jejímž cílem nebyla zdaleka jen
kvantitativní korekce kritických exponentù, ale která vnesla zcela nový pohled na øadu fyzikálních dìjù, se pøenáší zájem o popis kritických jevù na
složitìjší systémy, zahrnující napøíklad dynamické procesy, ale i jevy vyskytující se v tak rozlièných oborech, jako je napøíklad kosmologie,
biologie, ekonomie nebo sociologie.

Jedním z takových rozšíøení je studium \emph{povrchových fázových pøechodù}, které se odehrávají na rozhraní dvou nebo více fází. Pøedstavme si
pevnou podložku, na kterou rozlijeme jisté množství kapaliny. Ze zkušenosti víme, že mohou nastat dva pøípady: buï se kapalina rozprostøe po
podložce, nebo se snaží minimalizovat svùj kontakt s ní tak, že vytvoøí jednu èi více kapièek. To, jaký stav kapalina zaujme, lze dobøe popsat
známými makroskopickými zákony založených na bilanci sil pùsobících na spoleèném rozhraní všech tøí zúèastnìných fází: pevné stìny, kapaliny i
okolního plynu. Ovšem až na konci sedmdesátých let minulého století se ukázalo, že i tak zdánlivì jednoduchý proces jako je smáèení na rovinné
podložce s sebou nese celou øádu netriviálních aspektù, které makroskopické argumenty nejsou schopny postihnout. V první øadì se ukazuje, že zmìnou
vnìjších parametrù (teplota, tlak, interakèní potenciál podložky...) mùže kapalina pøejít z jednoho stavu do druhého, a tato pøemìna nese všechny
rysy fázového pøechodu. Znamená to, že v okolí fázového pøechodu vykazuje volná energie systému singulární chování, které se mùže projevit buï v
první derivaci volné energie, což odpovídá fázovému pøechodu prvního druhu, nebo ve vyšších derivacích a pak se jedná o spojitý fázový pøechod. Již
tento samotný fakt generuje celou øadu otázek: Jaký je øád fázového pøechodu smáèení? Jaký je vliv mikroskopických sil na charakter fázového
pøechodu? Jaký vliv mají tepelné fluktuace a dimenze systému? Jaké jsou v pøípadì spojitého pøechodu hodnoty kritických koeficientù a jsou jejich
hodnoty univerzální, nebo závisejí na povaze molekulárních interakcí? Existují ještì nìjaké další, \clqq pøidružené\crqq $ $ fázové pøechody
související se smáèením?  Jak se zmìní odpovìdi na všechny tyto otázky se zmìnou geometrie podložky?

Smyslem tohoto èlánku je pøedstavit alespoò zlomek z velmi zajímavé a spletité fenomenologie povrchových fázových pøechodù a  nastínit nìkteré
základní metody pro jejich popis. V tomto èlánku se zamìøíme na pøípad smáèení na rovinné, ideálnì hladké podložce a ukážeme si, že i v tomto
nejjednodušším pøípadì existuje pøekvapivì bohatá struktura jevù, z nichž nìkteré odolávaly uspokojivému vysvìtlení až do velmi nedávné minulosti.
Pøestože lze jejich popis (jak je zvykem pøedevším u poèítaèových simulací) formulovat pomocí diskrétních modelù Isingova typu a jednotlivé fáze
oznaèovat jako \clqq$+$\crqq $ $ nebo \clqq$-$\crqq, zde se omezíme na spojité modely popisující distribuci volnì se pohybujících èástic ve vnìjším
poli nemìnného pole pevné stìny, pøièemž vysokohustotní fázi budeme oznaèovat jako kapalinu a nízkohustotní fázi jako plyn. Z hlediska fázových
pøechodù jsou oba pøístupy ekvivalentní a, striktnì øeèeno, jejich øešení odpovídá nalezení partièní sumy/funkce pro daný model, což je obecnì
notoricky nároèný a jen málokdy proveditelný úkol statistické fyziky. Naštìstí se ukazuje, že pro popis povrchových jevù není vždy takto
mikroskopický pøístup nutný a že je možné získat takøka veškerou podstatnou informaci pomocí mesoskopických modelù, které z problému \clqq
vyintegrují\crqq $ $ nìkteré ménì dùležité stupnì volnosti. Výsledkem je tak úloha popisující efektivní interakci mezi dvìma fázovými rozhraními,
typicky stìna-kapalina a kapalina-plyn, jejichž prùmìrná vzdálenost urèuje stav systému a hraje roli parametru uspoøádání pro popis kritických jevù
smáèení. V druhé èásti se pak zmíním o nìkterých povrchových jevech, které se odehrávají na komplexnìjších strukturách než je ideálnì rovinná stìna,
kde vstupují do hry geometrické parametry stìny indukující nové typy fázových pøechodù a fluktuaèních režimù, které sice souvisí se smáèením, ale
jsou odlišné. U dostateènì strukturovaných modelù stìn je pak obzvl᚝ pùsobivá jemná souhra rùzných a vzájemnì se prolínajících typù povrchových
jevù, z nichž každý pøispívá do celkové mozaiky výsledného fázového chování na daném povrchu.


\section{Smáèení jako fázový pøechod}

Zaènìme následujícím myšlenkovým experimentem. Uvažujme nádobu s plynem opatøenou pístem, kterým mùžeme kontrolovat tlak uvnitø nádoby. Atomy plynu jsou pøitom v
kontaktu se stìnou nádobu, pøièemž pøedpokládáme, že atraktivní síly mezi atomy plynu a atomy stìny zpùsobují adsorpci velmi tenké, mikroskopické vrstvy kapaliny na
povrchu stìny. Stlaèujme nyní píst tak, že tlak $p$ nebo ekvivalentnì chemický potenciál $\mu$ systému roste ke své saturované hodnotì $p_0$ resp. $\mu_0$, odpovídající
rovnováze kapalina-pára v objemové fázi, tj. v nepøítomnosti vlivu stìn (dále již jen \clqq dvoufázové rovnováze\crqq). Pøedpokládáme pøitom, že celý dìj se odehrává za
konstantní podkritické teploty $T<T_c$. Ve dvoufázové rovnováze, formálnì v limitì $\mu\to\mu_0^-$, mohou nastat dvì situace: buï se na povrchu stìny vytvoøí
\emph{makroskopické} množství kapaliny, nebo nikoli. V prvním pøípadì mluvíme o stìnì, která je smáèena kapalinou; v tomto pøípadì hraje stìna roli nukleaèního centra,
na kterém se vytváøí nová fáze -- kapalina, která je v rovnováze stejnì stabilní jako její pára. Naopak, pokud šíøka kapalného filmu zùstane i v rovnováze pouze na
mikroskopické úrovni, pak kapalina stìnu nesmáèí a znamená to, že ke kondenzaci je zapotøebí jiného nukleaèního centra než je stìna (napø. zrnka prachu). Z hlediska
smáèení pøitom mùžeme stav stìny kvantifikovat pomocí povrchové hustoty nebo též adsorpce $\Gamma$, definované jako pøebytkové množství tekutiny, které je v systému
pøítomno v dùsledku pøítomnosti stìny na jednotku plochy a kterou definujeme jako
 \bb
 \Gamma(T,\mu)=\lim_{V,A\to\infty}\frac{N-\rho_g(T,\mu)V}{A}\,,\label{gamma}
 \ee
kde $N$ je celkový poèet atomù v systému (tedy v naší nádobì) o objemu $V$, $A$ je plocha stìny a $\rho_g$ odpovídá hustotì plynu v objemové fázi za daných podmínek. Z
výše uvedeného je zøejmé, že pro povrchové jevy, jako byl nᚠmyšlenkový experiment, je $\Gamma$ pøirozeným paramatrem uspoøádáním. Pro stìnu, která nesmáèí, je hodnota
tohoto parametru mikroskopická, a to i ve dvoufázové rovnováze, zatímco pro stìnu která je smáèena, bude $\Gamma$ v blízkosti rovnováhy divergovat (vliv gravitace
neuvažujeme).

Z pohledu teoretika indukuje popsaný experiment pøinejmenším dvì naléhavé otázky. Za prvé, proè a za jakých podmínek dochází k fázovému pøechodu smáèení a za druhé,
jakým zpùsobem lze tento spojitý fázový pøechod charakterizovat.

Abychom si odpovìdìli na první otázku, uvažujme opìt rozhraní stìna-plyn, na kterém se adsorbuje kapalný film prùmìrné šíøky $\ell_\pi$, jehož hodnota v blízkosti
rovnováhy souvisí s adsorpcí pøibližnì jako $\Gamma\approx\ell_\pi(\rho_l-\rho_g)$, kde $\rho_l$ je hustota saturované kapaliny. Pro danou hodnotu $\ell_\pi$ pøitom
existuje mnoho rozdílných konfigurací zvlnìného rozhraní kapalina-plyn, charakterizovaných  profilem lokální šíøky $\ell(\rr_\parallel)$, který závisí na souøadnicích
$\rr_\parallel$ paralelních se stìnou. Lokální šíøka filmu tak mùže být menší èi vìtší než $\ell_\pi$, pøièemž tyto fluktuace smìrem od stìny jsou omezeny pouze
podmínkou $\langle\ell(\rr_\parallel)\rangle=\ell_\pi$, zatímco fluktuace povrchu filmu smìrem ke stìnì jsou omezeny i pøítomností samotné (neprostupné) stìny. Z toho je
zøejmé, že s prùmìrnou šíøkou filmu $\ell_\pi$ roste poèet dovolených konfigurací, a tedy i entropie systému. Vidíme tedy, že entropie upøednostòuje co možná nejvyšší
hodnoty $\ell_\pi$, a ideálnì stav, kdy je stìna zcela smáèena. Lze tedy oèekávat, že ke smáèení stìny bude docházet spíše za vyšších teplot, kdy je vliv entropie
významný.

K tomu, zda stìna za dané teploty smáèena bude èi nikoli, má ovšem kromì entropie co øíct i energie systému. Její vliv mùžeme charakterizovat \emph{efektivním
potenciálem} $W(\ell)$, kterým pùsobí stìna na rozhraní kapalina-pára v závislosti na prùmìrné šíøce filmu. Odpovídá-li minimum této funkce koneèné hodnotì $\ell_\pi$,
pak bude energie preferovat \clqq vázaný stav\crqq, který zabraòuje úplnému smáèení stìny. V pøípadì, že globální minimum $W(\ell)$ je v nekoneènu, bude rozhraní filmu
od stìny odpuzováno a stìna smáèena bude. Jaká bude skuteèná rovnovážná šíøka filmu, bude nakonec záviset na bilanci obou faktorù, jak energie, tak entropie. Pokud
zanedbáme entropické vlivy a neuvažujeme fluktuace šíøky filmu okolo své støední hodnoty, redukuje se problém smáèení na aproximaci støedního pole spoèívající v prosté
minimalizaci $W(\ell)$. V opaèném pøípadì pøed námi stojí netriviální statisticko-mechanický problém, který spoèívá v nalezení partièní funkce modelového systému, a to
buï pøesnì (napø. pomocí metody matice pøechodu) nebo pøibližnì (pomocí teorie renormalizaèní grupy). To, do jaké míry je pro daný problém aproximace støedního pole
dostaèující, pøitom závisí na charakteru mezimolekulárních sil, na geometrii stìny (kterou zatím považujeme za rovinnou) a obecnì na dimenzi systému.

\begin{center}
\begin{figure}[h]
\includegraphics[width=8cm]{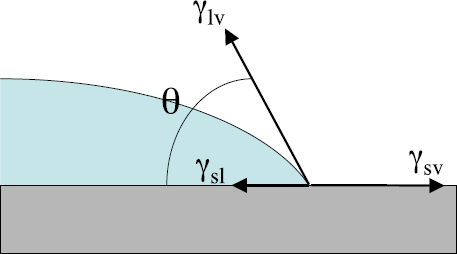}
\caption{Kapka s kontaktním úhlem $\theta>0$.  } \label{fig1}
\end{figure}
\end{center}

To jsme se ale posunuli ponìkud dál; vrame se na moment na zaèátek 19. století, do doby, kdy Thomas Young studoval kapilární jevy pomocí kontaktního úhlu
makroskopické kapky ležící na stìnì, viz. Obr.~1. Young ukázal, že kontaktní je urèen podmínkou rovnováhy
 \bb
 \gamma_{\rm sg}=\gamma_{\rm sl}+\gamma_{\rm lg}\cos\theta \label{young}
 \ee
pomocí povrchových napìtí na rozhraní stìna-plyn (sg), stìna-kapalina (sl) a kapalina-plyn (lg). Povrchové napìtí je èasto interpretováno mechanicky jako síla potøebná k
zvìtšení plochy povrchu. Pro nás ale bude užiteènìjší její termodynamická formulace. Pro systém o objemu $V$ a ploše mezifázového rozhraní $A$ mùžeme vyjádøit pøírùstek
energie pomocí fundamentální rovnice
 \bb
 dE=TdS+\mu dN-pdV+\gamma dA\,, \label{fund}
 \ee
kde $S$ je celková entropie systému a kde poslední èlen vyjadøuje práci potøebnou ke zvýšení plochy mezifázového rozhraní o $dA$. To, že funkce $E(S,N,V,A)$ je homogenní
øádu $1$ ve všech svých promìnných, nám umožòuje použitím Eulerova teorému  vyjádøit energii v integrálním tvaru:
 \bb
 E=TS+\mu N-pV+\gamma A\,. \label{Et}
 \ee

Pro popis fázových pøechodù je nejpøíhodnìjší  grand-kanonický soubor (fixujeme parametry $\mu$, $V$ a $T$), v nìmž roli volné energie hraje velký potenciál $\Omega$,
související s vnitøní energií dvojitou Legendreovou transformací
 \bb
 \Omega=E-TS-\mu N\,, \label{omega}
 \ee
jehož objemovou èást (nezahrnující pøíspìvek fázového rozhraní) lze vyjádøit jako $\Omega_b=-pV$.

Srovnáním (\ref{Et}) a (\ref{omega}) dostáváme, že povrchové napìtí lze též definovat jako povrchovou hustotu volné energie
 \bb
 \gamma=\lim_{V,A\to\infty}(\Omega-\Omega_b)/A\,.\label{omex}
 \ee

Lze ukázat (viz. napø. \cite{RW}), že pro diferenciál funkce $\gamma(T,\mu)$ platí
 \bb
 d\gamma=-S_sdT-\Gamma d\mu\,,\label{gibbs}
 \ee
 kde povrchová hustota entropie $S_s$ je definovaná analogicky k (\ref{gamma}) nebo  (\ref{omex}).


Vrame se k našemu popisu smáèení rovinné stìny pomocí šíøky kapalného filmu $\ell_\pi$ a ukažme, jak jej lze propojit s makroskopickou definicí smáèení založené na
Youngovì kontaktním úhlu. Pro rozhraní stìna-plyn je povrchovou hustotou volné energie $\gamma_{\rm sg}$. Je-li stìna smáèena, znamená to, že se na stìnì adsorbuje
kapalný film makroskopické šíøky. V tomto pøípadì je efektivní interakce mezi stìnou a rozhraním kapalina-plyn nulová (a mùžeme položit $W(\ell)=0$ pro $\ell\to\infty$),
takže cena, kterou systém musí zaplatit za adsorpci kapalného filmu spoèívá v pøítomnosti rozhraní stìna-kapalina, jejíž hodnota je $\gamma_{\rm sl}$ a dále za rozhraní
kapalina-plyn, jejíž výše je $\gamma_{\rm lg}$. Povrchovou energii na jednotku plochy odpovídající systému nasyceného plynu v pøítomnosti stìny je v tomto pøípadì tedy
 \bb
  \gamma_{\rm sg}=\gamma_{\rm sl}+\gamma_{\rm lg}\,.\label{wet}
 \ee
Pokud stìna smáèena není, je šíøka rozhraní koneèná a odpovídá minumu funkce $W(\ell)<W(\infty)$. Hodnota $W(\ell_\pi)$ je v tomto pøípadì záporná, a dostáváme tak
 \bb
  \gamma_{\rm sg}=\gamma_{\rm sl}+\gamma_{\rm lg}+W(\ell_\pi)<\gamma_{\rm sl}+\gamma_{\rm lg}\,.\label{nonwet}
 \ee
Srovnáním relací (\ref{wet}) a (\ref{nonwet}) s (\ref{young}) dospíváme k tomuto závìru: Je-li stìna smáèena, adsorbuje se na ní makroskopické množství kapalného filmu
$\ell_\pi\to\infty$, což odpovídá kontaktnímu úhlu $\theta=0$. Pokud stìna smáèena není, pak hodnota $\ell_\pi$ zùstává mikroskopická a kontaktní úhel je $\theta>0$. V
tomto pøípadì se makroskopické množství saturované kapaliny nanesené na stìnu rozpadne na malé kapièky s kontaktním úhlem daným rovnicí (\ref{young}). Takový stav stìny
se také nìkdy oznaèuje jako èásteèné smáèení.

\section{Fenomenologie smáèení a kritické koeficienty}

Z pøedchozích úvah vyplynulo, že to, zda bude stìna smáèena nebo ne, závisí významnì na teplotì. V roce 1977 publikuje J. Cahn zásadní èlánek \cite{cahn}, ve kterém
argumentuje, že jakmile se teplota systému zaène pøibližovat kritické teplotì kapalina-pára $T_c$, ke smáèení dojít musí. Jeho argument vychází z nerovnice
(\ref{nonwet}) a uvažuje teplotu blízkou $T_c$, pøi níž povrchové napìtí kapalina-pára klesá k nule jako
 \bb
 \gamma_{\rm lg}\sim(T_c-T)^{2\nu}\,,\label{nu}
 \ee
kde $\nu=0.65$ pro tøí-dimenzionální systém. Druhá èást jeho argumentace spoèívá v tom, že rozdíl $\gamma_{\rm sg}-\gamma_{\rm sl}$ je úmìrný rozdílu hustot kapaliny
a plynu vzhledem k tomu, že povrchová energie vzniká v dùsledku interakce atomù stìny s atomy plynu resp. kapaliny. V blízkosti kritické teploty tak platí
 \bb
\gamma_{\rm sg}-\gamma_{\rm sl}\propto(\rho_l-\rho_g)\sim(T_c-T)^\beta\,,\label{beta}
 \ee
kde $\beta\approx1/3$. Srovnání teplotních závislostí (\ref{nu}) a (\ref{beta}) ukazuje, že podmínka èásteèného smáèení (\ref{nonwet}) musí být pro teplotu dostateènì
blízkou k $T_c$ narušena, a tedy že stìna v blízkosti kritického bodu musí být smáèena.

Aèkoli detailnìjší analýza smáèení rovinné stìny v  blízkosti kritického bodu ukazuje, že i pøes svou pøesvìdèivost nejsou tyto makroskopické argumenty zcela pøesné,
znamenala Cahnova práce pøelom v teorii fázových pøechodù, kterou tak rozšíøila o popis povrchových (mezifázových) jevù. Její vývoj nebyl už od samého poèátku bez
kontroverze. Cahnova teorie pøedvídá, že  na køivce fázové rovnováhy existuje teplota smáèení $T_w<T_c$, pøi které dochází k fázovému pøechodu \emph{smáèení prvního
druhu}, kdy šíøka kapalného filmu (adsorpce) skoèí z koneèné hodnoty na hodnotu nekoneènou (makroskopickou). Tento jev byl skuteènì pomìrnì záhy ovìøen experimentálnì
\cite{moldover}. V té dobì však také publikuje Sullivan výsledky své elegantní mikroskopické teorie \cite{sullivan}, ze které vyplývá, že fázový pøechod v $T_w$ je
\emph{spojitý}; v tomto pøípadì na køivce fázové rovnováhy diverguje $\Gamma$ s teplotou spojitì pro $T\to T_w^-$. Tento jev se oznaèuje jako \emph{spojité} nebo
\emph{kritické smáèení}.

Stìna mùže být smáèena pouze tehdy, je-li adsorbovaná fáze (kapalina) termodynamicky stabilní, tedy pouze ve dvoufázové rovnováze ($\delta\mu \equiv\mu_0(T)-\mu=0$). Oba zmiòované fázové pøechody odpovídají termodynamickým
cestám, pøi kterých se pohybujeme po køivce fázové rovnováhy kapalina-pára a zvyšujeme teplotu až k $T_w$, kdy dojde ke smáèení. Druhý zpùsob jak mùže dojít ke smáèení odpovídá našemu myšlenkovému experimentu, pøi kterém se k
rovnováze pøibližujeme po izotermì, za pøedpokladu, že $T>T_w$. Tento fázový pøechod se nazývá \emph{úplné smáèení}.

Odkud se vlastnì bere naše oprávnìní oznaèovat tyto jevy za fázové pøechody? Fázový pøechod je jev, pøi kterém volná energie systému vykazuje singulární chování. To,
že tak tomu je ve všech tøech zmiòovaných pøípadech, je zjevné z toho, že dochází k divergenci $\Gamma$. Konkrétnì, úplné smáèení je spojitým fázovým pøechodem proto,
že první derivace volné energie diverguje pro $\mu=\mu_0(T>T_w)$, jak plyne z rovnice (\ref{gibbs}):
 \bb
 \frac{\partial\gamma}{\partial\mu}=-\Gamma\,.\label{Gamma}
 \ee
 Tuto divergenci charakterizujeme neuniversálním kritickým exponentem $\beta_{\rm co}$:
  \bb
  \ell\propto\Gamma\sim\delta\mu^{-\beta_{\rm co}}\,,
  \ee
jehož hodnota závisí na charakteru mezimolekulárních sil pøítomných v daném systému. Z hlediska povrchových jevù lze rozdìlit interakce do dvou tøíd: na dlouhodosahové
potenciály, které asymptoticky ubývají jako pøevrácená mocnina vzdálenosti a na krátkodosahové potenciály, které ubývají exponenciálnì nebo rychleji. Pro první (daleko
bìžnìjší) pøípad, lze ukázat, že v pøípadì kdy dominantní (tj. ubývající v nekoneènu nejpomaleji) pøíspìvek k celkové mezièásticové interakci má tvar $u\propto
r^{-(4+p)}$, je možné vyjádøit efektivní potenciál jako
 \bb
 W(\ell)=\delta\mu(\rho_l-\rho_g)\ell+\frac{A(T)}{\ell^p}+\frac{B(T)}{\ell^q}+\cdots\,,\label{W}
 \ee
kde typicky $q=p+1$ a speciálnì pro (neretardované) van der Waalsovy (disperzní) síly ve tøech dimenzích $p=2$. První èlen na pravé stranì zahrnuje termodynamickou cenu
za pøítomnost kapalné fáze, která je v objemové fázi nestabilní; je kladný pro úplného smáèení ($\mu<\mu_0(T)$) a nulový pro smáèení prvního druhu a kritické smáèení,
které probíhají za podmínky dvoufázové rovnováhy ($\mu=\mu_0(T)$). Další èleny rozvoje jsou dùsledkem mikroskopických interakcí mezi atomy stìny a tekutiny, pøièemž
koeficient u nejnižšího øádu $A(T)$ se nazývá Hamakerova konstanta.

Pro rovinnou stìnu zahrnuje efektivní potenciál veškerou závislost povrchové volné energie na šíøce kapalného filmu. Její hodnota odpovídá minimu funkce
 \bb
  \tilde{\gamma}_{\rm sg}(T,\delta\mu;\ell)=\gamma_{\rm sl}+\gamma_{\rm lg}+W(\ell)\label{cw}\,,
 \ee
takže $\gamma_{\rm sg}(T,\delta\mu)=\tilde{\gamma}_{\rm sg}(T,\delta\mu;\ell_\pi)$ \cite{note}.    Dosazením (\ref{W}) do (\ref{cw}) a minimalizací
$\tilde{\gamma}_{\rm sg}(T,\delta\mu;\ell)$ podle $\ell$ dostáváme
 \bb
 \ell_\pi\propto\Gamma\sim\delta\mu^{-\frac{1}{p+1}}\,,
 \ee
z èehož tak plyne $\beta_{\rm co}=1/(p+1)$ a speciálnì $\beta_{\rm co}=1/3$ pro disperzní síly. Ještì doplòme, že pro krátkodosahové potenciály $\beta_{\rm co}=0(\ln)$,
a tedy $\ell_\pi\sim-\ln(\delta\mu)$.

Chování singulárního pøíspìvku k povrchovému napìtí $\gamma_{\rm sg}$ pro úplné smáèení charakterizuje kritický exponent $\alpha_{\rm co}$,
 \bb
 f_{\rm sing}(T,\delta\mu)\equiv\gamma_{\rm sg}-\gamma_{\rm sl}-\gamma_{\rm lg}\sim\delta\mu^{2-\alpha_{\rm co}}\,.
 \ee
Z relace (\ref{Gamma}) ovšem vyplývá, že hodnoty kritických koeficientù $\alpha_{\rm co}$ a $\beta_{\rm co}$ nejsou nezávislé, nýbrž jsou vázány vztahem
 \bb
 \beta_{\rm co}=\alpha_{\rm co}-1\,.
 \ee

\begin{center}
\begin{figure}[h]
\includegraphics[width=10cm]{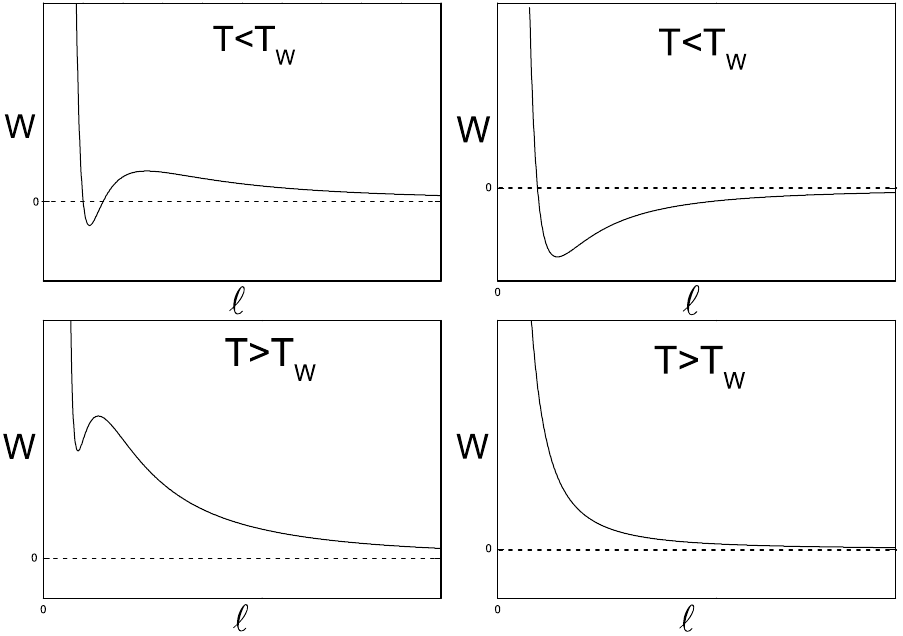}
\caption{Efektivní potenciál pro pøípad fázového pøechodu smáèení prvního druhu (vlevo) a kritického (spojitého) smáèení (vpravo). Systémy vykazující fázový pøechod
prvního druhu se vyznaèují pøítomností energetické bariéry mezi dvìma potenciálovými minimy, jejichž hodnoty se rovnají pøi teplotì $T=T_w$. Pro kritické smáèení je
charakteristická pøítomnost jednoho minima, jehož poloha se spojitì posouvá do nekoneèných hodnot a vymizí pro $T\to T_w$.} \label{W_sketch}
\end{figure}
\end{center}

V pøípadì procesu kritického smáèení roste šíøka adsorbovaného filmu spojitì pro $T\to T_w$ a tuto divergenci charakterizujeme pomocí kritického koeficientu
$\beta_s$:
 \bb
 \ell_\pi\sim t^{\beta_s}\,,
 \ee
kde jsme zavedli oznaèení $t\equiv(T_w-T)/T_w$. Je zøejmé, že pro $T>T_w$, kdy je stìna smáèena, musí efektivní potenciál klesat k nule shora, a tudíž speciálnì pro
(\ref{W}) je $A(T>T_w)$ kladné. Naopak pro $T<T_w$ existuje \clqq vázaný stav\crqq $ $ odpovídající minimu $W(\ell)$ pro koneèné $\ell$, jehož poloha se s rostoucí
teplotou spojitì posouvá k nekoneènu a $A(T<T_w)$ je záporné (viz. Obr.~\ref{W_sketch}, vpravo). Z toho plyne, že ke kritickému smáèení dojde právì když $A=0$ (a $B>0$),
a tedy že $A\sim t$ (v tìchto úvahách je patrná analogie s Landauovou teorií kritického bodu).

Pro urèení hodnoty $\beta_s$ pro disperzní síly postupujeme podobnì jako v pøípadì úplného smáèení a minimalizujeme (\ref{cw}), pøièemž klademe $\delta\mu=0$.
Dostáváme
 \bb
 \ell_\pi^{q-p}(T,\delta\mu=0)=-\frac{qB}{pA}\,,\;\;\;T<T_w\,,
 \ee
což pro experimentálnì nejrelevantnìjší pøípad van der Waalsových sil ve tøech dimenzích $p=2$, $q=3$ dává
 \bb
 \ell_\pi(T,\delta\mu=0)=-\frac{3B}{2A}\,,\;\;\;T<T_w\,,
 \ee
a tedy
 \bb
 \ell_\pi\sim t^{-1}\Rightarrow \beta_s=-1\,,
 \ee
 (obecnì $\beta_s=-1/(q-p)$). Pro krátkodosahové potenciály je divergence opìt logaritmická $\ell_\pi(T,\delta\mu=0)\sim-\ln|t|$, a tedy $\beta_s=0(\ln)$.

Kritický exponent charakterizující singulární chování povrchového napìtí pro kritické smáèení
 \bb
 f_{\rm sing}(T,\delta\mu=0)\sim t^{2-\alpha_s}\,.\label{fsing}
 \ee
je roven $\alpha_s=-1$, jak vyplývá z dosazení pøedchozích výsledkù do (\ref{cw}) a kde jsme položili $p=2$. Odsud také mùžeme okamžitì dostat teplotní závislost
kontaktního úhlu. Srovnáním (\ref{young}),  (\ref{nonwet}) a (\ref{fsing}) dostáváme vztah mezi exponentem $\alpha_s$ a kontaktním úhlem
 \bb
 1-\cos\theta\sim t^{2-\alpha_s}\,, \label{theta_alpha}
 \ee
 z èehož pro $\alpha_s=-1$ a $t\to0$ kdy $\theta\to0$ plyne
  \bb
  \theta\sim t^{3/2}\,.
  \ee
Vidíme tak, že v pøípadì kritického smáèení klesá kontaktní úhel v blízkosti teploty smáèení k nule nejen spojitì, ale navíc hladce.

Mechanismus smáèení je odlišný, jedná-li se o fázový pøechod prvního druhu. Tento pøípad je charakterizován pøítomností energetické bariéry u efektivního potenciálu a
soutìžením dvou minim funkce $W(\ell)$ (viz. Obr.~\ref{W_sketch}, vlevo). Zatímco v pøípadì kritického smáèení se minimum funkce $W(\ell)$ posouvá smìrem k nekoneèným
hodnotám až pro $T=T_w$ vymizí, v tomto pøípadì minimum funkce $W(\ell)$ pro koneèné $\ell$ roste s rostoucí teplotou od záporných hodnot (kdy je minimem globálním) k
nule pro $T=T_w$, kde se vyrovná limitní hodnotì $W(\ell\to\infty)$. Pro vyšší teploty pøedstavuje toto minimum pro koneèné $\ell$ pouze metastabilní stav. Je zøejmé, že
tvar efektivního potenciálu vyžaduje nyní zahrnutí minimálnì ještì jednoho èlenu vyššího øádu v (\ref{W}) (pøièemž $A<0$ a $B>0$), a je tedy zapotøebí uvažovat i úèinek
krátkodosahových pøíspìvkù mezimolekulárních sil. Protože se jedná o fázový pøechod prvního druhu, musí dojít k výmìnì latentního tepla, a tedy ke skoku povrchové
entropie
 \bb
 S_s=-\frac{\partial\gamma_{\rm sg}}{\partial T}\,,
  \ee
a tedy k nespojitosti v derivaci kontaktního úhlu podle teploty v bodì $t=0$. Vzhledem k tomu, že pro $t\to0^+$ vyplývá z (\ref{theta_alpha})
  \bb
  \frac{d\cos\theta}{dT}\sim t^{1-\alpha_s}\,,\;t\to0^+\,,
  \ee
urèuje podmínka nespojitosti hodnotu $\alpha_s=1$, takže pøi smáèení prvního druhu ubývá kontaktní úhel  pro $t\to0^+$ ($T\to T_w^-$) jako  $\theta\sim t^{1/2}$. Z
charakteru chování kontaktního úhlu v blízkosti teploty smáèení tak lze experimentálnì celkem snadno rozhodnout, o jaký øád fázového pøechodu se jedná.


\begin{center}
\begin{figure}[h]
\includegraphics[width=7cm]{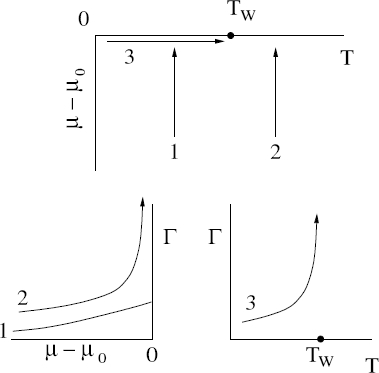}  \hspace*{1cm}  \includegraphics[width=7cm]{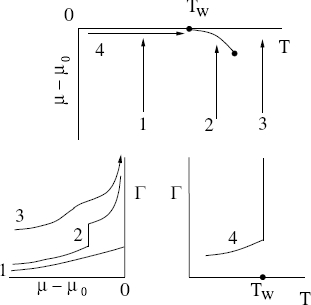}
\caption{{\bf Vlevo}: Povrchový fázový diagram znázoròující pøípad {\bf kritického smáèení}. Nad teplotou $T_w$ je stìna smáèena, pod teplotou $T_w$
je smáèena pouze èásteènì. Na diagramu jsou znázornìny tøi termodynamické cesty. Cesty $1$ a $2$ znázoròují izotermy, podél nichž roste chemický
potenciál smìrem k dvoufázové rovnováze. Pro cestu $2$, kdy $T>T_w$, dochází v blízkosti rovnováhy k úplnému smáèení. Cesta $3$ sleduje dvoufázovou
rovnováhu a její limita $T\to T_w^-$ odpovídá kritickému smáèení. Chování adsorpce pro daný proces je zobrazen v grafech dole. {\bf Vpravo}:
Povrchový fázový diagram znázoròující pøípad {\bf smáèení prvního druhu}. Nad teplotou $T_w$ je stìna smáèena, pod teplotou $T_w$ je smáèena pouze
èásteènì. Na diagramu jsou znázornìny ètyøi termodynamické cesty. Cesty $1$, $2$ a $3$ znázoròují izotermy, podél nichž roste chemický potenciál
smìrem k dvoufázové rovnováze. Pro cesty $2$ a $3$, kdy $T>T_w$, dochází v blízkosti rovnováhy k úplnému smáèení. Podél cesty $2$ dochází navíc ke
skoku v adsorpci o koneènou hodnotu v bodì protnutí køivky pøedsmáèení. Tento skok se zmenšuje s rostoucí teplotou (a se vzdáleností od rovnováhy),
až vymizí zcela v kritickém bodì pøedsmáèení. K fázovému pøechodu smáèení prvního druhu dojde pøi teplotì $T_w$  podél cesty $4$. Chování adsorpce
pro daný proces je zobrazen v grafech dole. } \label{fig3}
\end{figure}
\end{center}

\begin{center}
\begin{figure}[h]
\includegraphics[width=5cm]{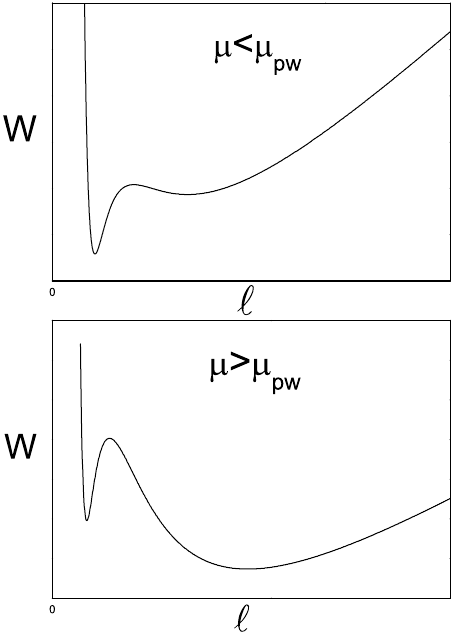}
\caption{Ilustrativní prùbìh efektivního potenciálu znázoròující fázový pøechod pøedsmáèení. Mimo dvoufázovou rovnováhu ($\delta\mu=0$) nemùže být
stìna zcela smáèena díky pøítomnosti prvního èlenu v (\ref{W}). Srovnání s prùbìhem funkce $W(\ell)$ pro pøípad smáèení prvního druhu
(Obr.~\ref{W_sketch}, vlevo) ukazuje, že pøídání lineárního èlenu v $\ell$ posune druhé minumum do koneèné vzdálenosti. Obì minima mají stejnou
hodnotu pro $\mu=\mu_{\rm pw}$, kdy dojde  ke skoku v šíøce filmu o koneènou hodnotu, a na stìnì tak mohou koexistovat dva filmy o rùzných
(koneèných) šíøkách. Situace odpovídá cestì 2 na Obr.~3, vpravo.} \label{fig_prewet}
\end{figure}
\end{center}

Podobnì jako fázové diagramy objemových systémù zobrazují množinu bodù, v nichž má pro dané hodnoty termodynamických parametrù volná energie
singularitu, znázoròují \emph{povrchové fázové diagramy} množinu bodù, kde vykazuje singulární chování povrchová volná energie. Na obrázku 3 jsou
schematicky tyto diagramy zobrazeny pro kritické smáèení (vlevo) a smáèení prvního druhu (vpravo). Pro fázový pøechod prvního druhu je
charakteristická ještì pøítomnost køivky zobrazující fázový pøechod \emph{pøedsmáèení}. Tato køivka, která je prodloužením singularity volné energie
v $T_w$ do vyšších teplot a pro $\delta\mu>0$ odpovídá skoku v adsorpci bìhem procesu úplného smáèení o koneènou hodnotu, která se s teplotou
postupnì zmenšuje, až vymizí v kritickém bodì pøedsmáèení $T_{\rm cpw}$, který spadá do tøídy univerzality dvoudimenzionálního Isingova modelu. Oba
faktory, totiž že køivka pøedsmáèení je krátká i to, že se ke køivce rovnováhy $\delta\mu=0$ napojuje teènì, jsou dùvody, proè se na experimentální
ovìøení existence tohoto fázového pøechodu èekalo až do devadesátých let \cite{prewet}, aèkoli k jeho pøedpovìzení došlo  již v ranných stádiích
teorie smáèení \cite{cahn, ebner}. Prùbìh efektivního potenciálu pro tento proces je znázornìn na obrázku \ref{fig_prewet}.

\section{Vliv fluktuací, korelaèní funkce}

Až dosud byly všechny naše úvahy na úrovni teorie støedního pole, která uvažuje pouze nejpravdìpodobnìjší stav systému.  Vliv fluktuací  je možné
zahrnout pomocí modelu efektivního hamiltoniánu
 \bb
 H[\ell]=\int \dd{\bf x} \left\{\frac{\gamma_{\rm lg}}{2}(\nabla\ell({\bf x}))^2+W(\ell({\bf x}))\right\} \label{heff}\,,
 \ee
kde ${\bf x}$ je $d-1$-dimenzionální vektor paralelní se stìnou ($d$ je dimenze systému). V rámci tohoto modelu uvažujeme  odchylky
$\delta\ell\equiv\ell({\bf x})-\ell_\pi$ lokální výšky rozhraní od jeho støední hodnoty $\ell_\pi=\langle\ell\rangle$.
První èlen v hamiltoniánu zahrnuje tepelnými fluktuacemi generovaný pøíspìvek  odpovídající (entropicky výhodnému a
energeticky nevýhodnému) zvrásnìní rozhraní kapalina-pára, které vede k nárùstu její plochy o faktor
$\sqrt{1+(\nabla\ell)^2}\approx\frac{1}{2}(\nabla\ell)^2$; druhý èlen pak popisuje efektivní interakci mezi tímto rozhraním a stìnou. Na úrovni
støedního pole, kdy zanadbáváme vliv fluktuací, je $\nabla\ell({\bf x})\equiv0$ (pro rovinnou stìnu), takže rovnovážný stav systému je dán prostou
minimalizací energetického pøíspìvku k $H[\ell]$, tedy podmínkou $W'(\ell)=0$ s øešením $\ell({\bf x})=\ell_\pi$, což vede na výše uvedené výsledky.
Vliv fluktuací lze v prvním pøiblížení studovat pomocí funkcionálního rozvoje (\ref{heff}) do druhého øádu okolo  extrému $H[ell_\pi]$:




 \bb
H[\ell_\pi+\delta\ell]\approx H[\ell_\pi]+\frac{1}{2}\int\dd{\bf x}\left\{\gamma_{\rm lg}\nabla^2+W''(\ell_\pi)\right\}(\delta\ell({\bf x}))^2\,.
\label{gauss}\,,
 \ee
 Vyjádøíme-li tyto fluktuace ve Fourierovì obraze, $\delta\ell(\bf x)=\sum_{\bf k}{\rm e}^{i{\rm k}\cdot{\rm
x}}\delta\tilde\ell(\bf k)$, dostáváme pro druhý èlen v (\ref{gauss}) výraz
 \bb
 H_G[\ell]=\frac{1}{2V}\sum_{\bf k}\left\{\gamma_{\rm lg}k^2+W''(\ell_\pi)\right\}|\delta\tilde\ell({\bf k})|^2\,, \label{k_gauss}
 \ee
který popisuje vliv fluktuací na úrovni \emph{gaussovské aproximace}. Odsud  použitím ekvipartièního teorému okamžitì vyplývá:
  \bb
  \langle|\delta\tilde\ell({\bf k})|^2\rangle=\frac{k_BTV}{\gamma_{\rm lg}}\frac{1}{\xi_\parallel^{-2}+k^2}\,, \label{lk}
  \ee
   kde jsme zavedli velièinu s rozmìrem délky
  \bb
\xi_\parallel\equiv\left(\frac{\gamma_{\rm lg}}{W''(\ell_\pi)}\right)^\frac{1}{2}\,,\label{xi_par}
 \ee
  jejíž význam vyplyne za chvíli.

Korelaci mezi fluktuacemi v lokální výšce rozhraní popisuje korelaèní funkce
 \bb
 G(x_{12})\equiv\langle\delta\ell({\bf x_1})\delta\ell({\bf x_2})\rangle=\langle\delta\ell({\bf 0})\delta\ell({\bf x_{12}})\rangle\,,\label{G}
 \ee
kde jsme využili translaèní symetrie podél roviny stìny. Vzhledem k tomu, že $\langle|\delta\tilde\ell({\bf k})|^2\rangle=V\tilde{G}({\bf k})$,
dostáváme s použitím (\ref{lk})
  \bb
  \tilde{G}(k)=\frac{k_BT}{\gamma_{\rm lg}}\frac{1}{\xi_\parallel^{-2}+k^2}\,, \label{gk}
  \ee
a tedy
   \bb
  G(x)=\frac{k_BT}{\gamma_{\rm lg}}\frac{1}{(2\pi)^{d-1}}\int\frac{\dd^{d-1} k\, {\rm e}^{i{\bf k}\cdot{\bf x}}}{\xi_\parallel^{-2}+k^2}\,, \label{gx}
  \ee
kde implicitnì pøedpokládáme, že velikost vlnového vektoru $k$ je omezena vysokofrekvenèním \clqq cut-offem\crqq $ $ $\Lambda$.  Lorentzovský tvar
Fourierova obrazu korelaèní funkce je dùsledkem aproximace (\ref{gauss}) a je charakteristický pro \emph{Ornsteinovu-Zernikeho (OZ) teorii}
korelaèních funkcí.

Dùležitou vlastností korelaèní funkce je její škálovací charakter. Transformací $ k\to k \xi$ mùžeme $G(x)$ vyjádøit ve formì
 \bb
 G(x)=\frac{1}{x^{d-3}}g(x/\xi_\parallel)\,.\label{G_asym}
 \ee
 pøièemž z asymptotického tvaru $G(x)$ ($x\gg\xi_\parallel$):
 \bb
 G(x)\sim x^{(2-d)/2} {\rm e}^{-x/\xi_\parallel}\,,
 \ee
lze interpretovat $\xi_\parallel$ jako \emph{paralelní korelaèní délku},  která tak pøedstavuje pøirozenou délkovou škálu pro korelaèní funkci.  Z
definièního vztahu (\ref{xi_par}) je zøejmé, že paralelní korelaèní délka diverguje pro $t\to0$ a tuto divergenci  lze charakterizovat pomocí
kritického exponentu $\nu_\parallel$ definovaném relací $\xi_\parallel\sim t^{-\nu_\parallel}$, pøièemž $\nu_\parallel=(q+2)/2(q-p)$ pro
dlouhodosahové potenciály (\ref{W}).

Výsledky OZ teorie, a speciálnì tvar korelaèní funkce (\ref{G_asym}), jsou pro povrchové jevy enormnì dùležité.  Tak napøíklad  tzv. sumaèní pravidlo
(nìkdy též oznaèováno jako fluktuaèní-disipaèní teorém, nebo se jedná o vztah mezi korelaèní a responzivní funkcí) vyplývající z rovnic (\ref{gk}) a
(\ref{xi_par}) dává do vztahu \clqq susceptibilitu\crqq $\chi^{-1}=\beta\partial^2f_s/\partial\ell^2$ a korelaèní funkci:
 \bb
 \chi=\tilde{G}(0)\propto\int G(x) \dd^{d-1}x=\int \dd^{d-1}x\frac{1}{x^{d-3}}g(x/\xi_\parallel)=\int \dd^{d-1}y\frac{\xi_\parallel^2}{y^{d-3}}g(y)
 \propto\xi_\parallel^2\,.
 \ee
kde jsme použili škálovací relaci(\ref{G_asym}). Pro $t\to0$ $\chi$ zjevnì diverguje a tuto divergenci lze  charakterizovat zákonem $\chi\sim
t^{-\gamma}$ definujícím kritický exponent $\gamma$. Z poslední rovnice pak okamžitì dostáváme
 \bb
 \gamma=2\nu_\parallel\,.
 \ee

Analogické výsledky vyplývající z OZ teorie dostáváme i pro objemové systémy (dimenze $d-1$ je nahrazena $d$).  Zde ovšem nemá pøesné (tedy po
zahrnutí vlivu fluktuací) kritické chování korelaèní funkce tvar analogický k (\ref{G_asym}), ale
 \bb
 G(x)=\frac{1}{x^{d-2-\eta}}g(x/\xi)\,,\label{G_asym2}
 \ee
z èehož také vyplývá
 \bb
\gamma=(2-\eta)\nu\,. \label{eta}
 \ee
 Pøítomnost nového (nenulového) exponentu $\eta$ souvisejícího s tzv. anomální dimenzí  má hluboké opodstatnìní. Pøipomeòme si, že OZ teorie
bere v úvahu pouze dlouhovlnné fluktuace, tedy takové, kdy se parametr uspoøádání mìní pomalu; vyšší než kvadratické pøíspìvky gradientu parametru uspoøádání zanedbává.
Tyto dlouhovlnné fluktuace mùžeme charakterizovat korelaèní délkou $\xi$ (objemovou analogií k $\xi_\parallel$), kterou OZ teorie definuje a která v kritickém bodì
diverguje s koeficientem $\nu$. Kdyby korelaèní délka byla jedinou relevantní délkovou škálou v kritické oblasti, dospìli bychom dimenzionální analýzou k tomu, že
$\eta=0$, a tedy že OZ teorie je pøesná! To by ovšem vedlo k paradoxu, protože OZ teorie je stále na úrovni teorie støedního pole, jejíž výsledky jsou v rozporu s
experimentem. Východiskem z tohoto problému je pøítomnost ještì jiné délkové škály, která musí být pøesnou teorií zahrnuta. Bude-li vliv fluktuací významný, mùže
napøíklad v plynné fázi kromì plynulé zmìny hustoty docházet k nukleaci zárodkù kapalné fáze. Jejich rozmìr, který rozhodnì není nijak popsán OZ teorií, vstupuje do dìje
jako druhá a dokonce významnìjší velièina s rozmìrem délky. Je-li zahrnuta, dostáváme nenulovou hodnotu $\eta$, která tak mìøí \clqq míru nepøesnosti\crqq  $ $ OZ
teorie.

Výbornou zprávou je, že pro povrchové jevy je tvar korelaèní funkce (\ref{G_asym})  pøesný i v pøípadì, kdy je vliv fluktuací významný, a tedy že  $\eta=0$
\cite{lip_fish}. Analogií nukleace nové fáze je právì smáèení na stìnì, kdy šíøka adsorbovaného filmu diverguje. Tato divergence je doprovázena fluktuacemi, které jsou
zcela lokalizovány na rozhraní filmu a jiné fluktuace související se vznikem dalších fluktuaèních center se již nevyskytují (pøedpokládáme, že nejsme v tìsné blízkosti
kritické teploty $T_c$). Tyto povrchové fluktuace, tzv. \emph{kapilární vlny},  jsou OZ teorií ovšem popsány velmi dobøe, nebo odpovídají dlouhovlnným módùm této teorie
a mají za následek postupné odpoutání rozhraní filmu od stìny. Význam OZ teorie pro povrchové jevy je tak ještì vìtší než pro systémy objemové.

\section{Škálovací režimy}

Zvrásnìní rozhraní v dùsledku tepelných fluktuací mùžeme kvantifikovat pomocí \emph{hrubosti}, jejíž kvadrát definujeme jako
$\xi_\perp^2\equiv\langle\delta\ell^2\rangle$. Z definice je zøejmé, že $\xi_\perp^2=G(0)$, a tedy že
 \begin{eqnarray}
  \xi_\perp^2&=&\frac{k_BT}{\gamma_{\rm lg}}\frac{1}{(2\pi)^{d-1}}\int\frac{\dd^{d-1} k}{\xi_\parallel^{-2}+k^2}\,, \label{xiperp}\\
             &\propto&\int_0^\Lambda\frac{\dd k\,k^{d-2}}{\xi_\parallel^{-2}+k^2}\,.\label{xiperp2}
  \end{eqnarray}
U tohoto výrazu se na chvilku zastavíme. Charakter integrálu zjevnì závisí na dimenzi a lze jej rozdìlit na tøi pøípady:
 \bb
 \xi_\perp\sim\left\{\begin{array}{cc}\Lambda^{d-3}&d>3\\
                         (\ln(\Lambda \xi_\parallel))^\frac{1}{2}&d=3\\
                         \xi_\parallel^{(3-d)/2}&d<3
                         \end{array}\right. \label{wander}
 \ee
Vidíme tedy, že pøítomnost  $\xi_\parallel$ regularizuje integrál v (\ref{xiperp2}) vùèi infraèervené divergenci  (vùèi ultrafialové divergenci je integrál regularizován
uvažováním \clqq cut-offu\crqq $ $ $\Lambda$; ten odpovídá pøevrácené hodnotì nejmenší délkové škály, což pro diskrétní systémy Isingova typu je møížková konstanta a pro
spojité systémy je to atomová vzdálenost  nebo korelaèní délka  $\xi$). V nepøítomnosti stìny (kdy $\xi_\parallel=0$) by integrál ve výrazu pro hrubost (\ref{xiperp2})
divergoval pro relevantní dimenze $d\leq3$ a taktéž by divergovala korelaèní funkce $G(x)$ pro $x\to\infty$. To znamená, že pro tyto dimenze je volné rozhraní
kapalina-pára \emph{hrubé} v tom smyslu, že pokud bychom znali polohu tohoto rozhraní v nìjakém daném bodì, nejsme schopni prohlásit nic o poloze rozhraní v jiném, hodnì
vzdáleném bodì. To souvisí s tím, že pro volné rozhraní je energie nutná pro jeho posunutí nulová, jak je také vidìt z výrazu (\ref{k_gauss}) pro $k\to0$. Jinými slovy,
mód $k=0$ je pro volné rozhraní \emph{Goldstoneùv mód}. Z tìchto úvah také vyplývá obecná dùležitost dlouhovlnných fluktuací (kapilárních vln) pro jevy na fázových
rozhraní.

\begin{center}
\begin{figure}[h]
\includegraphics[width=5cm]{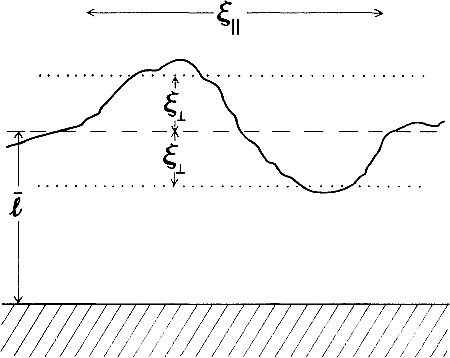}
\caption{Ilustrace fluktuujícího rozhraní kapalina-pára vázaného k rovinné stìnì.} \label{xi}
\end{figure}
\end{center}

Naopak rozhraní, které je vázáno ke stìnì ($\ell_\pi$ je koneèné) nikdy hrubé ve výše uvedeném smyslu není. Výchylky výšky rozhraní mohou být znaèné, ale vždy koneèné.
Omezená je i korelaèní funkce, což znamená, že známe-li výšku rozhraní $\ell({\bf x})$ v jednom bodì, pak víme, že poloha rozhraní v jiném, libovolnì vzdáleném bodì
spadá pøibližnì do intervalu  $\ell({\bf x})\pm\xi_\perp$. Formálnì lze rozdíl mezi volným a vázaným rozhraním vyjádøit pøítomností efektivního nízkofrekvenèního \clqq
cut-offu\crqq $ $ $\xi_\parallel^{-1}$ v integrálech pro korelaèní funkci nebo hrubost:
 \bb
 \int_0^\Lambda \dd k k^{d-4}\to\int_{{\xi_\parallel}^{-1}}^\Lambda \dd k k^{d-4}\,.
 \ee
U kritického ($t\to0$) a úplného smáèení ($\delta\mu\to0$) paralelní korelaèní délka diverguje jako  $\xi_\parallel\sim t^{-\nu_\parallel}$ (kritické smáèení)  a
$\xi_\parallel\sim \delta\mu^{-\nu_{\parallel}^{\rm co}}$ (úplné smáèení), což definuje kritické koeficienty $\nu_\parallel$ a $\nu_{\parallel}^{\rm co}$. V dùsledku
této divergence dochází i k divergenci $\xi_\perp$, což je v konzistenci s pøedchozími závìry, nebo v tomto pøípadì se rozhraní odpoutané od stìny stává volným.


Platnost OZ teorie je omezena oprávnìností ukonèení funkcionálního rozvoje (\ref{gauss}) druhým øádem. Z této aproximace je zøejmé, že OZ teorie nezahrnuje celé
spektrum fluktuací, ale pouze dlouhovlnné pøíspìvky, a nelze tedy pøedpokládat její obecnou platnost. Výsledky z ní plynoucí jsou stále na úrovni aproximace støedního
pole, její význam je však pøesto mimoøádný. Jednak nám umožòuje nahlédnout do mikroskopické struktury tekutiny, a to skrze experimentálnì mìøitelné velièiny
\cite{OZ}, a jednak lze do jisté míry využít i k identifikaci režimù, v nichž samotná OZ teorie již neplatí.

To, za jakých okolností je pøítomnost fluktuací významná, lze odhadnout následujícím zpùsobem. V úvodu jsme si øekli, že pokud jsou fluktuace natolik velké, že rozhraní
èasto \clqq naráží do stìny\crqq, a tedy $\xi_\perp\sim\langle\ell\rangle$, bude to znamenat pokles entropie systému, oproti volnému rozhraní (bez pøítomnosti stìny).
Každá srážka rozhraní se stìnou \clqq stojí\crqq  $ $  øádovì $k_B$, a tedy ztráta (povrchová) hustoty entropie v dùsledku pøítomnosti stìny je øádovì $\Delta
s\approx-k_b/\xi_\parallel^{d-1}$. Pøítomnost fluktuací má dále za následek energeticky nevýhodné zvlnìní rozhraní, jehož \clqq cenu\crqq  $ $ lze odhadnout (viz.
Obr.~\ref{xi}) z prvního èlenu v (\ref{heff}) jako $ \Delta e \approx \gamma_{\rm lg}(\xi_\perp/\xi_\parallel)^2$. Z (\ref{wander}) plyne, že speciálnì pro $d<3$ (a jak
vyplyne z dalšího, jen v tomto pøípadì mùže být pro dalekodosahové interakce vliv fluktuací významný) jsou energetické a entropické pøíspìvky fluktuací k volné energii
stejného øádu
 \bb
\Delta f_s=\Delta e -T\Delta s\sim\xi_\parallel^{1-d}\sim\xi_\perp^{2(d-1)/(d-3)}\,.\label{ff}
  \ee
Jsou-li fluktuace významné $\xi_\perp\sim\langle\ell\rangle$, a jejich efekt lze zahrnout prostým pøidáním dalšího interakèního èlenu do efektivního potenciálu. V
pøípadì kritického smáèení ($\delta\mu=0$) víme, že v rámci teorie støedního pole jsou podstatné první dva netriviální èleny rozvoje (\ref{W}), takže
 \bb
 W_{\rm fl}(\ell)=A(T)\ell^{-p}+B(T)\ell^{-q}+C(T)\ell^{-\tau}+\cdots \label{Wfl}
 \ee
 kde $\tau=2(d-1)/(3-d)$ a typicky $q=p+1$. Relativní významnost tohoto efektivního fluktuaèního èlenu urèuje odpovídající \emph{fluktuaèní režim}:

 \begin{itemize}

\item $\tau>q$: {\bf Režim støedního pole (MF)}. V tomto pøípadì je fluktuaèní èlen nepodstatný, a jeho pøítomnost tak nemá na výsledky plynoucí z minimalizace
                     (\ref{W}) žádný vliv. Speciálnì,  divergence šíøky filmu $\ell_\pi\sim t^{-\beta_s}$ pro $T\to T_w$ je charakterizována kritickým koeficientem
                     $\beta_s=1/(q-p)$.

\item  $q\geq\tau> p$: {\bf Režim slabých fluktuací (WFL)}. V tomto režimu je fluktuaèní èlen druhý nejvìtší a nepodstatným se stává èlen øádu $\ell^{-q}$. Teplota
smáèení, kterou urèuje první èlen, se tak oproti pøedpovìdím teorie støedního pole nemìní, hodnoty kritických koeficientù ovšem ano, a to tak, že ve všech výrazech pro
kritické exponenty $\tau$ nahradí $q$, takže napøíklad $\beta_s=1/(\tau-p)$. Znamená to, že na rozdíl od klasických kritických exponentù (vyplývajících z teorie
støedního pole), jsou v tomto režimu kritické exponenty závislé na dimenzi prostoru.

\item  $\tau\leq p$: {\bf Režim silných fluktuací (SFL)}. V tomto režimu je chování rozhraní již naprosto dominováno fluktuacemi. V tomto pøípadì již nelze pouze
\clqq opravit\crqq $ $ výsledky plynoucí z teorie støedního pole, ale je tøeba použít zcela jiné metody pro øešení partièní funkce $Z=\int{\cal{D}}\ell e^{-H[\ell]}$. V
mnoha pøípadech lze nalézt pøesná øešení pomocí metody matice pøechodu, což vede na øešení diferenciální rovnice Schr\"{o}dingerova typu pro pravdìpodobnost realizace
stavu s daným rozhraním $\ell({\bf x})$. Pro aproximativní øešení lze použít metody funkcionální renormalizaèní grupy \cite{rg}.

 \end{itemize}

Pro dimenzi $d=3$ je efektivní fluktuaèní èlen exponenciálnì malý, jak vyplývá z (\ref{wander}), a nemùže tak konkurovat algebraicky ubýhajícím èlenùm. V tomto
pøípadì existuje pouze MF režim a jediný vliv fluktuací je zanedbatelnì malá hrubost rozhraní $\xi_\perp\approx\sqrt{\ln\ell_\pi}$. Zanedbatelný efekt fluktuací pro
tøídimenzionální systémy taktéž vyplývá z toho, že podmínka $q=\tau$ urèuje mezní, tedy horní kritickou dimenzi pro kritické smáèení:
 \bb
 d^*=\frac{3q+2}{q+2}\,,\label{dcrit}
 \ee
z níž vyplývá $d^*<3$ pro libovolné koneèné $q$, což je velmi dùležitý výsledek. Dostáváme tedy, že pro fyzikálnì nejrelevantnìjší tøídu tøídimenzionálních systémù s
(neretardovanými) van der Waalsovými silami $d^*=11/5$, a tedy že pøedpovìdi teorie støedního pole jsou pro tuto tøídu systémù pøesné. Situace je složitìjší pro systémy
s krátkodosahovými silami pro nìž $q=\infty$, nebo z (\ref{dcrit}) plyne, že v tomto pøípadì $d^*=3$.

Význam fluktuací pro úplné smáèení ($\delta\mu>0$) je menší, pøièemž existují pouze dva fluktuaèní režimy. To vyplývá z tvaru efektivního potenciálu
 \bb
 W_{\rm fl}(\ell)=\delta\mu\ell+A(T)\ell^{-p}+C(T)\ell^{-\tau}+\cdots\,,\label{Wflcw}
 \ee
kde opìt $\tau=2(d-1)/(3-d)$. Oproti pøípadu kritického smáèení, není èlen øádu $\ell^{-q}$ relevantní ani v režimu støedního pole, zato èlen lineární v $\ell$ je
relevantní vždy, a neexistuje tak SFL. Zcela formálnì bychom mohli použít efektivního potenciálu ve tvaru (\ref{Wfl}) s tím, že koeficient $p=-1$ (a odpovídající faktor
$a=\delta\mu$); relevance tohoto èlenu (a tedy nepøítomnost SFL) je pak zøejmá, nebo $\tau>0$. Možné režimy tedy jsou:

 \begin{itemize}

\item $\tau>p$: {\bf Režim støedního pole (MF)}. V tomto pøípadì je vliv fluktuací zanedbatelný, takže výsledky teorie støedního pole jsou pøesné. Platí tedy
napøíklad $\beta^{\rm co}=1/(p+1)$, $\alpha^{\rm co}=(p+2)/(p+1)$ nebo $\nu_\parallel^{\rm co}=(2+p)/2(p+1)$.

\item  $\tau\leq p$: {\bf Režim slabých fluktuací (WFL)}. V tomto pøípadì dominují fluktuace a všechny kritické exponenty závisí pouze na $d$, a tedy napøíklad
 $\beta^{\rm co}=(3-d)/(1+d)$, $\alpha^{\rm co}=4/(1+d)$ nebo $\nu_\parallel^{\rm co}=2/(1+d)$.

 \end{itemize}

Horní kritickou dimenzi urèuje tentokrát podmínka $p=\tau$, a tedy
 \bb
 d^*_{\rm co}=\frac{3p+2}{p+2}\;\;\;\,,\label{dco}
 \ee
Vidíme tedy, že i pro úplné smáèení je horní kritická dimenze $d^*_{\rm co}<3$ pro dlouhodosahové potenciály a $d^*_{\rm co}=3$ pro krátkodosahové potenciály. Speciálnì,
pro systémy s van der Waalsovými silami je pro $d=3$ koeficient v (\ref{Wflcw}) $p=2$, takže $d^*_{\rm co}=2$ a fluktuace nejsou tedy podstatné. Ovšem ve dvou dimenzích
je tento exponent roven $p=3$, a tedy $d^*_{\rm co}=11/5>2$ a fluktuace jsou již dominantní.

Srovnejme tyto výsledky se standardním kritickým chováním objemového systému. Pøipomeòme si, že v tomto pøípadì máme dvì škálovací pole, teplotu a
napøíklad chemický potenciál, a obì musí nabýt konkrétních hodnot, aby bylo dosaženo kritického bodu. To znamená, že z celé množiny všech kritických
koeficientù jsou pouze dva nezávislé a ostatní lze z nich odvodit z relací jako je napø. (\ref{eta}).  Za tyto dva nezávislé koeficienty se typicky
bere $\nu$ a $\eta$. V pøípadì kritického smáèení máme taktéž dvì relevantní škálovací pole: kritický bod je dán podmínkami $\mu=\mu_0$ a $T=T_w$.
Jak ovšem víme, na rozdíl od standardního kritického bodu je nyní $\eta=0$, a existuje tak jediný nezávislý kritický koeficient (napø.
$\nu_\parallel$), jehož hodnota urèuje i všechny ostatní. Pokud pøedpokládáme, že je vliv fluktuací významný, dostáváme napøíklad z (\ref{ff})
\emph{hyperškálovací relaci}
 \bb
 2-\alpha=(d-1)\nu_\parallel\,\;\;\;d\leq d^*\,,\label{hyper}
 \ee
z které (známe-li $\nu_\parallel$) získáme singulární chování volné energie, které nám umožní urèit další kritické koeficienty. Tyto závìry souvisí s výše uvedenými
výsledky tak, že (pro danou dimenzi) lze všechny kritické koeficienty vyjádøit pomocí jediného parametru $p$. Jak je to v pøípadì úplného smáèení? Úplné smáèení je
jednodušší než kritické smáèení, nebo zde figuruje pouze jediné škálovací pole, $\mu$ (podobnì jako u fázových pøechodù prvního druhu objemových systémù); teplota mùže
nabývat libovolných hodnot v intervalu $(T_w,T_c)$. Oproti kritickému smáèení tak máme ještì o jeden nezávislý parametr ménì, a všechny kritické koeficienty jsou tedy
vázany pøímo na dimenzi systému.

Zpùsob, jakým jsme urèili hodnoty horních kritických dimenzí pro kritické a úplné smáèení, vyžadoval výpoèet koeficientu $\tau(d)$ rozmìrovou analýzou. Je velmi
ilustrativní ukázat, jak se horní kritické dimenze dají stanovit i bez této znalosti, a to pomocí škálovací metody, která je zárodkem teorie renormalizaèní grupy, ale
jejíž výpoèetní nároky jsou minimální. Zaènìme pøípadem kritického smáèení. Víme, že pro dimenze $d\leq d^*$ je relevantní pouze první (netriviální) èlen rozvoje
(\ref{W}), takže staèí uvažovat hamiltonián ve tvaru
 \bb
 H[\ell]=\int \dd^{d-1}x \left\{\frac{\gamma_{\rm lg}}{2}(\nabla\ell({\bf x}))^2+\tilde{A}t\ell^{-p}\right\}\,,
 \ee
kde jsme využili toho, že $A(t)\sim t$ a definovali koeficient $\tilde{A}$, který je koneèný pro $t\to0$. Uvažujme nyní škálovací transformaci
  \begin{eqnarray}
   X&=&xt^a\nonumber\\
   L&=&\ell t^b\,.\label{scale_t}
   \end{eqnarray}
Škálovací exponenty $a$ a $b$ zvolíme tak, aby transformovaný hamiltonián již nezávisel na $t$, a nabyl tak formy
 \bb
 H[L]=\int \dd^{d-1}X \left\{\frac{\gamma_{\rm lg}}{2}(\nabla L({\bf X}))^2+\tilde{A}L^{-p}\right\}\,, \label{HL}
 \ee
což vede k podmínkám $a=(2+3p-pd)/2(d-1)$ a $b=a(3-d)/2$. Nezávisí-li (\ref{HL}) na $t$, nemùže ovšem na $t$ záviset ani støední hodnota transformované šíøky filmu, tzn.
$\langle L\rangle={\cal{O}}(1)$ je èíslo, a musí tedy platit
 \bb
 \ell_\pi\sim t^{-b}\,.
 \ee
 Zároveò víme, že pro dimenze $d\geq d^*$ platí výsledek z teorie støedního pole $\ell_\pi\sim t^{-\frac{1}{q-p}}$. Tyto dva výsledky se musí shodovat právì pro $d=d^*$,
 z èehož vyplývá $d^*=(3q+2)/(q+2)$, v souladu s (\ref{dcrit}).

 Podobnì postupujeme pro úplné smáèení. Pro dimenze $d\leq d^*$ je nyní relevantní pouze první èlen rozvoje (\ref{W}), takže uvažujeme hamiltonián ve tvaru
  \bb
 H[\ell]=\int \dd^{d-1}x \left\{\frac{\gamma_{\rm lg}}{2}(\nabla\ell({\bf x}))^2+\delta\mu\ell\right\}\,,\label{Hco}
 \ee
 který vyjádøíme v pøeškálovaných promìnných
  \begin{eqnarray}
   X&=&x\delta\mu^a\nonumber\\
   L&=&\ell\delta\mu^b\,,\label{scale2}
   \end{eqnarray}
tak aby (\ref{Hco}) již nezáviselo na $\delta\mu$, tedy
 \bb
 H[L]=\int \dd^{d-1}X \left\{\frac{\gamma_{\rm lg}}{2}(\nabla L({\bf X}))^2+L\right\}\,. \label{Hco2}
 \ee
 Z podmínek $a=2/(d+1)$ a $b=(3-d)/(d+1)$ a známého výsledku z teorie støedního pole $\ell_\pi\sim\delta\mu^{-\frac{1}{p+1}}$ dostáváme (\ref{dco}).

\vspace{0.5cm}

Ukázali jsme si, že teoretické modely pøedvídají singulární chování volné energie pøi smáèení na rovinné stìnì, pøièemž charakter tìchto singularit závisí na dosahu
mikroskopických interakcí, dimenzi systému a termodynamické cestì. Co se týèe experimentálního ovìøení tìchto predikcí, nejjednoznaènìji se ukazuje pøípad úplného
smáèení, kde mìøení plnì potvrzují teoretické výsledky \cite{comw}. Cesta $T\to T_w$ podél dvoufázové rovnováhy je experimentálnì nároènìjší. Z hlediska øádu fázového
pøechodu se ukazuje, že možné jsou oba pøípady, pøièemž obecnì platí, že èím blíže je teplota smáèení $T_w$ kritické teplotì $T_c$, tím vyšší je pravdìpodobnost, že
smáèení bude spojité (kritické). Z èetných mìøení dále vyplývá, že smáèení prvního druhu je v pøírodì mnohem bìžnìjší. To vyplývá z toho, že kritické smáèení klade
podstatnì restriktivnìjší podmínky na parametry mezimolekulárních interakcí. První experiment prokazující existenci kritického smáèení byl realizován teprve ke konci
minulého století \cite{ross}, a to pouze pro pøípad, kdy úlohu stìny hraje jiná kapalina. Pro pevnou stìnu nebylo kritické smáèení detegováno dodnes.

\end{document}